# Adsorption Barrier Limits the Ice Inhibition Activity of Glycan-Rich Antifreeze Glycoproteins

*Wentao Yang, Zhaoru Sun**

School of Physical Science and Technology, ShanghaiTech University, Shanghai 201210, China

**ABSTRACT:** Antifreeze glycoproteins (AFGPs) are among the most potent ice recrystallization inhibition (IRI) agents, yet the molecular basis for their counterintuitive decline in activity with increasing glycosylated threonine (T*) content remains unresolved. Through molecular dynamics simulations of model glycoproteins with increasing T* content, we show that the potent IRI activity of AFGPs arises not only from the thermodynamic stability of strong ice-binding states, but also from their kinetic accessibility. Specifically, the free energy barrier for forming strong ice-binding states from the unbound state constitutes a critical kinetic bottleneck. Increasing T* content enhances the overall hydration capacity due to the additional glycan moieties, thereby imposing a greater desolvation penalty and elevating the adsorption barrier. This kinetic limitation, rather than the absence of strong ice-binding states, accounts for the experimentally observed decline in IRI activity. To quantify the structural basis of this behavior, we introduce a facial amphiphilicity index that integrates both spatial segregation and compositional ratio of hydrophilic and hydrophobic residues, and show that it correlates well with IRI activity. These findings highlight that facial amphiphilicity mediates a critical balance between binding stability and kinetic accessibility, providing a rational design principle for advanced IRI materials.

## Introduction

Cryoprotectants are essential for preserving cell viability during cryopreservation by effectively inhibiting ice recrystallization (IR), thereby preventing osmotic imbalance and mechanical damage induced by large ice crystals.[1-3] Consequently, ice recrystallization inhibition (IRI) activity has emerged as a central criterion in the development of advanced cryoprotectants.[4-6] Over recent decades, a wide range of natural and synthetic IRI-active materials—such as antifreeze proteins (AFPs),[7,8] antifreeze glycoproteins (AFGPs),[9,10] polymers,[11-15] small molecules,[16-18] and supramolecular assemblies[19-21]—have been extensively investigated. Among them, AFGPs, typically consisting of repeating tripeptide units (Ala-Thr-Ala)$_n$ in which the hydroxyl group of the Thr is glycosylated with the disaccharide (Galβ1-3GalNAcα1-), remain one of the most potent IRI agents known.[22] However, their high cost,[23] massive production issues,[24] and potential immunogenicity[25] impede their practical applications in both medical and industrial settings. A comprehensive understanding of their structure-activity relationship is paramount for the rational design of next-generation IRI agents.

It is widely accepted that AFGPs function via an adsorption-inhibition mechanism in which the proteins bind to the ice surfaces and prevent ice growth.[26,27] Within this framework, considerable efforts have been devoted to identifying the structural motifs responsible for ice binding.[28,29] Chemical modification experiments have shown that both methyl groups and disaccharide hydroxyl groups are indispensable for the potent IRI activity of AFGPs, as their removal leads to a significant decrease in activity.[30,31] Molecular dynamics (MD) simulations further revealed that these groups can directly bind to ice,[32-34] and their high abundance in AFGPs underpins strong ice-binding stability, contributing to the exceptional IRI activity.[35,36] While these findings highlight the importance of binding-site number, recent evidence suggests that this view may be oversimplified.[37,38] For example, Kramer et al.[38] recently reported that synthetic AFGPs with increasing glycosylated Thr (T*) content (~33%, ~50%, and ~67%) exhibit reduced IRI activity, despite which induces more hydroxyl and methyl groups. These observations imply that the number of ice-binding sites alone cannot fully account for ice binding stability and IRI efficiency, underscoring the need for a more detailed mechanistic understanding of AFGP-ice interactions at the atomic level.

In addition to the binding stability, an increasing number of studies emphasized the critical role of adsorption kinetics in inhibiting ice growth.[39-42] Meister et al.[26] used fluorescence microscopy to demonstrate that the extraordinary IRI activity of AFGPs arises from their rapid adsorption onto ice surfaces. Similarly, Thosar et al.[43] recently showed that the binding and accumulation of AFPs on ice is limited by the kinetics of adsorption. Further support comes from the simulation study of Naullage and Molinero,[44] who revealed that the impaired ice-binding ability of acetylated poly(vinyl alcohol) does not stem from destabilization of the bound state, but rather from large kinetic barriers preventing the propagation of binding. These studies collectively highlight that adsorption and accumulation are inherently time-dependent processes, with rapid adsorption during ice growth being essential for effective inhibition.[45] This is particularly relevant for IRI activity, where reversible binding is sufficient to suppress recrystallization.[39,46] Nevertheless, how increasing T* content affects the adsorption kinetics of AFGPs on the ice surface at the atomic level remains poorly understood.[38] A comprehensive investigation of both thermodynamic and kinetic aspects of the binding is therefore essential to elucidate the molecular mechanisms underlying their T* content-dependent IRI activity.

In this work, we employ all-atom MD simulations to elucidate the molecular origin of the IRI activity in glycoproteins with increasing T* content, using three model sequences (Figure 1a): AAT*AAT*AAT*AAT*AA (AAT), AT*AT*AT*AT*AT*AT*AA (ATAT), and AT*T*AT*T*AT*T*AT*T*AA (ATT). Firstly, replica-exchange MD simulations show that all three proteins adopt dominant conformations with polyproline II (PPII)-like backbone structures, yet exhibit distinct glycan orientations. Our simulations reveal an ice inhibition trend of AAT > ATAT > ATT, content with experimental observations.[38] Further analyses demonstrate that, beyond binding stability, the adsorption barrier—defined as the free energy required to transition from the unbound state to the strong binding state—critically governs adsorption kinetics, and thereby dictating the overall IRI efficacy. Excessive T* content increases the desolvation penalty due to the disruption of hydrogen bonds during adsorption, thereby raising the adsorption barrier and hindering stable binding. To quantify the structural basis of these effects, we propose a facial amphiphilicity index that integrates both the ratio and spatial segregation of hydrophilic and hydrophobic groups. This index correlates well with IRI performance, highlighting the importance of facial amphiphilicity in optimizing both binding stability and kinetic accessibility.

## Methods

**Molecular Dynamics Simulation.** All MD simulations are performed with GROMACS 2020.7[47] using the CHARMM36 force field[48,49] along with the TIP4P/Ice water model.[50] Lorentz-Berthelot combination rules are used to describe the cross-interaction parameters. The cutoffs for the van der Waals and Coulombic interactions are set to 1.2 nm, and long-range electrostatic interactions are evaluated with the particle-mesh Ewald (PME) algorithm.[51] The LINCS algorithm[52] is employed to constrain the covalent chemical bonds including hydrogen atoms. The leapfrog algorithm is used to numerically integrate the equation of motion with a time step of 2 fs. The temperature T and pressure P for all production runs (including enhanced sampling simulations) are controlled using the Nosé-Hoover thermostat[53,54] and the Parrinello-Rahman barostat[55] with time constants of 0.5 and 2.0 ps, respectively. The pressure is set to 1 atm in all NPT-MD simulations. Periodic boundary conditions are used in three directions.

**Replica Exchange MD Simulations.** Replica exchange MD (REMD) simulations[56] are performed to enhance the conformational sampling for AAT, ATAT, ATT in solution. The proteins with a conformation that separates the hydrophilic and hydrophobic parts are placed in a dodecahedral box with 4750 water molecules. The system is first energy-minimized using the steepest descent algorithm. Subsequently, it undergoes a two-step equilibration protocol: 500 ps in the NVT ensemble followed by 200 ps in the NPT ensemble at 300 K, controlled by the Nosé-Hoover thermostat[55] and the Berendsen barostat,[57] respectively. The resulting configuration is then used as the starting point for REMD simulations. Forty-eight replicas are employed at temperatures ranging from 268 to 400 K, generated using the algorithm of van der Spoel.[58] Each replica was further equilibrated for 500 ps in the NPT ensemble with the Berendsen barostat and Nosé-Hoover thermostat. Finally, production REMD simulations are carried out for 400 ns per replica in the NPT ensemble, with temperature exchanges attempted every 10 ps between adjacent replicas. The average exchange ratios between nearest neighbor temperatures are 27.3±5.7%, 27.1±5.7% and 27.1±5.7% for AAT, ATAT, and ATT, respectively.

**Ice inhibition simulations.** At the initiation of the simulation, the protein with PPII structure is positioned ~1.0 nm above the ice prismatic plane consists of four layers of proton disordered hexagonal ice generated from the Genice program,[59] and then solvated with 11000 water molecules in the liquid phase. The dimensions of the simulation box are 6.3 × 6.6 × 12.0 $nm^3$. The oxygen atoms within the ice slab are positionally constrained using a harmonic potential with a force constant of 1000 kJ $mol^{-1}$ $nm^{-2}$. Three independent simulations are conducted for each protein. A 500 ps NVT-MD run at 268 K is performed for equilibration after energy minimization. Final production NVT-MD run is evolved for 800 ns at 268 K to monitor the growth of ice.

The CHILL+ algorithm,[60] which has been shown to provide reliable accuracy in identifying ice molecules from liquid water,[32,61] is employed to classify water molecules into either ice or liquid phases and to monitor the evolution of the ice-water interface throughout the simulations. To characterize the binding behavior of the protein above the dynamically advancing ice front, the protein-ice distance was calculated as the average vertical (z-direction) distance between all heavy atoms of the protein and their nearest oxygen atoms directly beneath them on the ice surface. In addition, the orientation of the protein relative to the ice surface was described by the cosine of the angle between the protein backbone—defined by the line connecting the Cβ atoms of the first and last alanine residues—and the z-axis. Both the protein-ice distance and protein orientation are evaluated over the production trajectories (20-750 ns) to obtain statistically representative results.

**Metadynamics simulations.** The well-tempered metadynamics (WTMetaD),[62] a powerful and well-established enhanced sampling technique that drives the system to explore the entire free energy surface with respect to selected collective variables (CVs), was employed to explore probe the ice affinity of different proteins. All metadynamics simulations are carried out using PLUMED 2.8.[63,64]

The simulation systems consist of a six-layer proton-disordered hexagonal ice slab, a single protein molecule, and approximately 9,000 water molecules. The temperature was maintained at 280 K to prevent ice growth. To suppress melting, the ice molecules in the bottom five layers were harmonically restrained, while in the topmost layer, only the oxygen atoms were restrained. This allowed molecular rotation and thus better represented realistic ice growth dynamics. Prior to the WTMetaD simulation, a 500 ps NVT-MD run at 280 K is performed for equilibration after energy minimization. During the WTMetaD simulations, two CVs are employed to bias the sampling: (i) the protein-ice distance (d, CV1), defined as the projection along the z-axis between the protein's center of mass and the mean z-coordinate of the oxygen atoms in the uppermost ice layer of the ice template, and (ii) the cosine of the angle ($\cos\theta$, CV2) between the protein backbone and the z-axis, as also described in the Ice Inhibition Simulations section. The WTMetaD parameters are set as follows: Gaussian width ($\sigma$) was 0.02 nm and 0.05 for CV1 and CV2, and the bias factor ($\gamma$) was set to 50. Gaussians with an initial height (W) of 1.0 kJ/mol were deposited every 500 MD steps. Each WTMetaD simulation is conducted for more than 2.0 μs in the NVT ensemble, and terminated only when the reconstructed free energy surface exhibits fluctuations smaller than 0.5 kJ/mol within the last 50 ns, indicating convergence. The final free energy profiles were calculated using the reweighing technique proposed by Tiwary and Parrinello.[65]

## Results and Discussion

We begin our study by examining the conformational characteristics of the proteins (Figure 1a) with varying T* contents in aqueous solution using REMD simulations. To assess their secondary structure, we construct Ramachandran plots (Figure 1b), where the free energy is expressed as a function of the backbone dihedral angles derived from the $\phi$ and $\psi$ dihedral angle distributions $P(\phi,\psi)$ : $F(\phi,\psi) =- k_B T ln[P(\phi,\psi)]$ .[32,66] Notably, all three proteins predominantly adopt PPII helix structure (> 70%), with its population increases with T* content (AAT < ATAT < ATT), in agreement with previous experimental findings.[38] However, this opposes the experimentally observed decline in IRI activity (AAT > ATAT > ATT),[38] suggesting that considering the secondary (PPII) structure alone cannot account for their distinct IRI efficiency, even though it has been suggested to play a key role in the function of AFGPs.[32,66] Therefore, further investigation into the three-dimensional conformational features is warranted to elucidate the structural basis underlying the IRI activity of these proteins.

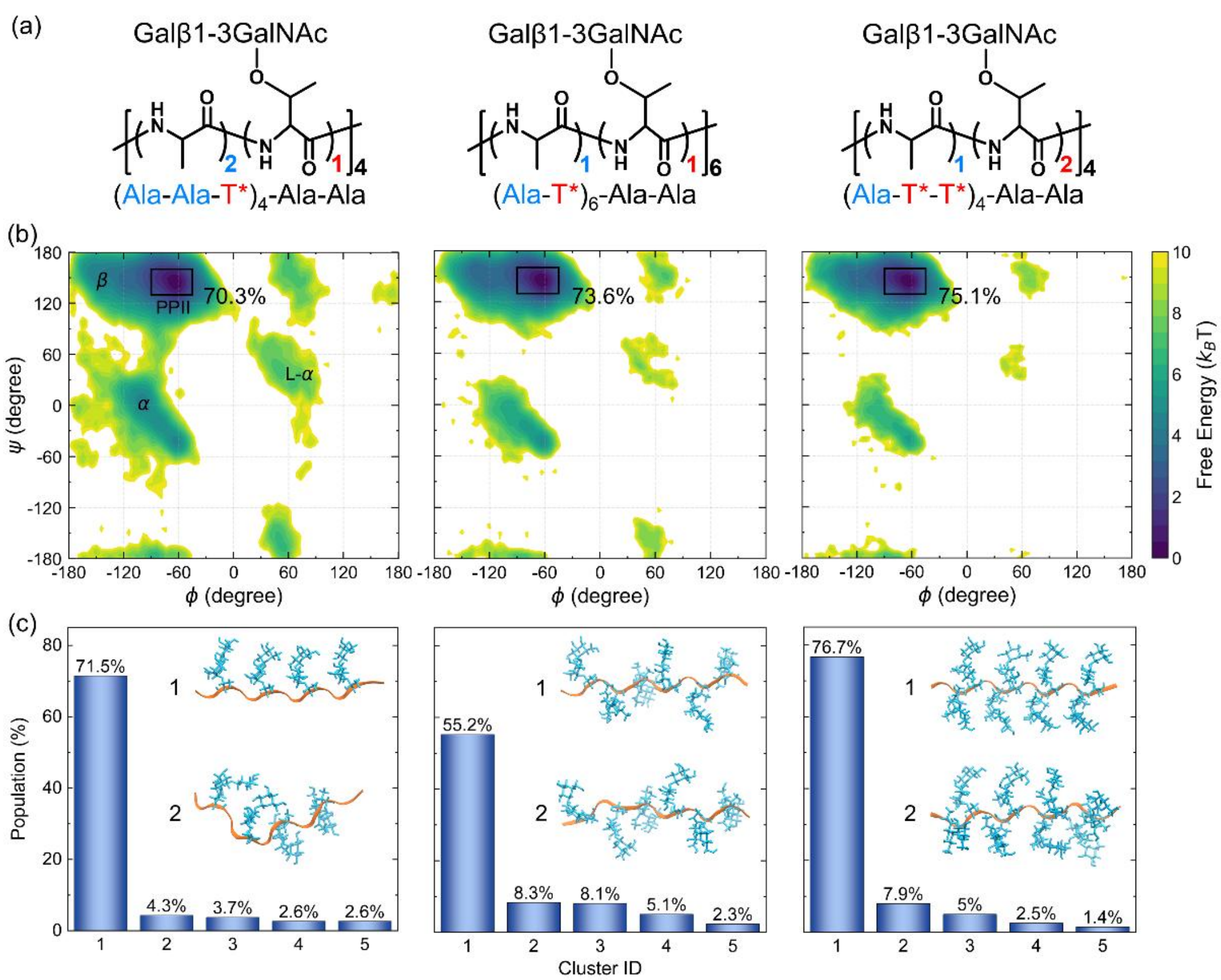

Figure 1. Structures and conformations of proteins with increasing glycosylated Thr content (a) Chemical structures of AAT (left), ATAT (middle), and ATT (right), illustrating increasing

numbers of glycosylated Thr (T*). Glycan moieties are represented as Galβ1-3GalNAc, where Gal is β-D-galactose and GalNAc is α-N-acetyl-D-galactosamine. (b) Ramachandran plots of the proteins at 268 K, with the PPII helix region indicated by black rectangles. (c) Histogram of the top five conformational cluster populations from REMD simulations. Representative three-dimensional structures of the two most populated clusters are shown for each peptide; backbones are displayed in orange (NewRibbon), and T* residues are displayed in cyan (Licorice). In all panels, data for AAT, ATAT, and ATT are presented from left to right.

To explore their three-dimensional structures, we then performed clustering analysis of the REMD trajectories using the Gromos Clustering Algorithm[67] with a root-mean-square deviation (RMSD) cutoff of 3.5 Å for all heavy atoms. This approach enabled us to identify distinct conformational states and evaluate their relative stabilities and population distributions. As shown in Figure 1c, each protein exhibits a well-defined dominant conformational cluster, with the top cluster accounting for over 55% of the total population. This demonstrates that the proteins adopt highly stable conformations in solution, with their dominant conformers likely playing a critical role in their distinct ice inhibition efficiencies.[68] Accordingly, all subsequent simulations were performed on these dominant conformers.

Moreover, representative structures from the top two clusters for each protein are also illustrated in Figure 1c. For AAT, the dominant conformation exhibits a clear spatial segregation of hydrophobic and hydrophilic regions, with all glycan chains aligned on one side of the backbone, in agreement with previous simulations.[32,69] In contrast, ATAT displays a less defined glycan orientation and lacks a clear amphiphilic pattern, which may underlie its reduced IRI activity.[68] Notably, ATT adopts a highly glycan orientation ordered V-shaped conformation (side view), in which glycans are quasi-symmetrically distributed on both sides of the backbone (Figure 1c), resulting in an apparent facial amphiphilicity. However, ATT has been reported to exhibit the lowest IRI activity,[38] suggesting that conformational order and amphiphilic pattern alone are insufficient to account for the observed differences in activity. We hypothesize that the ice growth inhibition ability of these proteins is directly governed by their binding mechanisms at the ice-water interface.

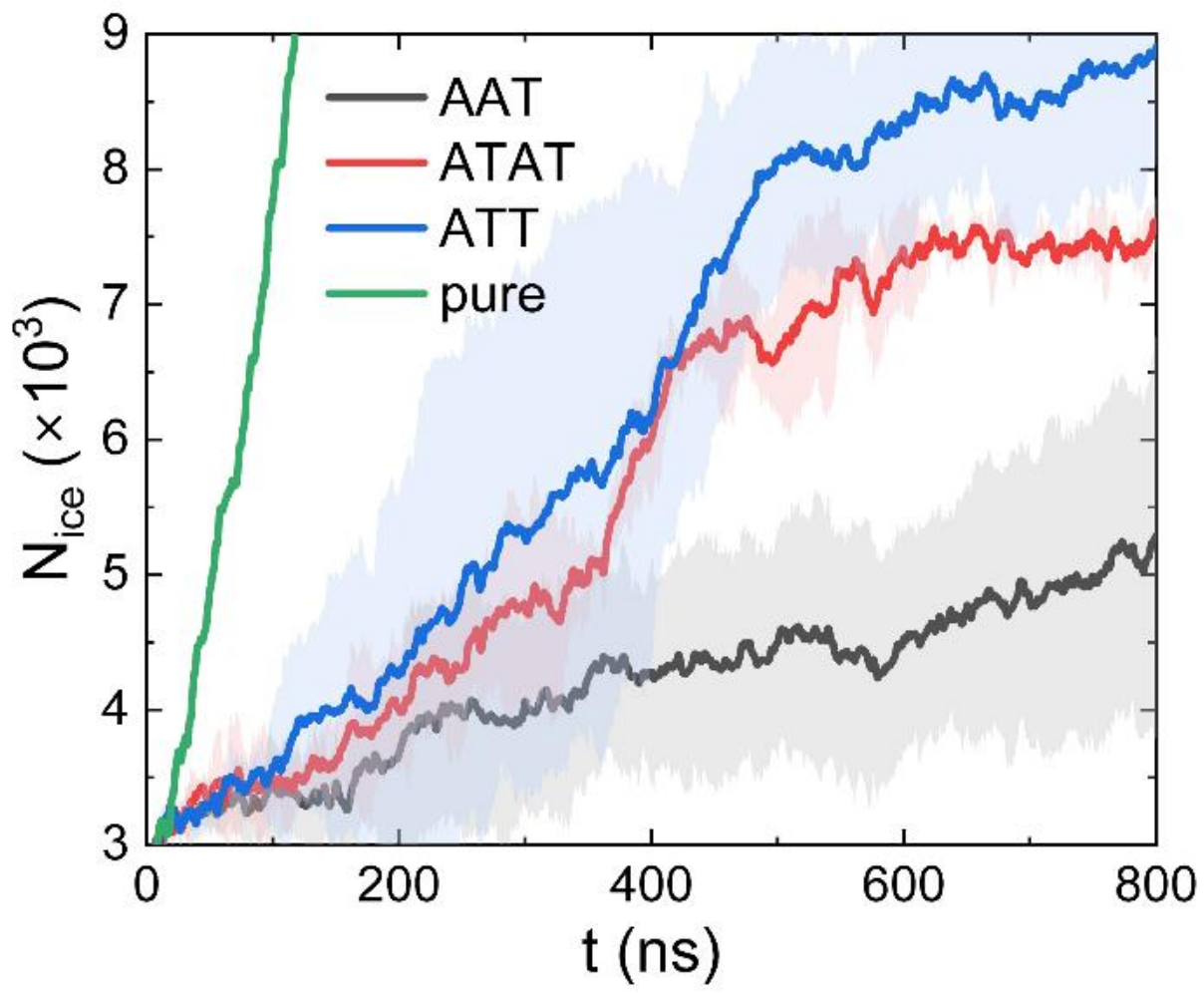

Figure 2. Ice growth inhibition behaviors of AAT (black), ATAT (red), ATT (blue), and the protein-free control (green). Data represent averages from three independent simulations, with error bars indicated as shaded regions.

To evaluate the inhibition ability of these proteins on ice growth, we conducted simulations of a protein-ice-water system using their dominant conformations. This set up, which has been widely adopted for computational antifreeze studies,[32,33,61,68,70-72] enables direct evaluation of protein-ice interactions and their influence on ice growth (see Section Methods for details). As shown in Figure 2, the number of ice molecules over time was monitored as a quantitative measure of ice inhibition ability. AAT consistently exhibited strong suppression of ice growth throughout the entire simulation. In contrast, both ATAT and ATT showed weaker inhibition during the initial 500 ns, with ice growth proceeding more rapidly in this early stage. Notably, ATAT enters the deceleration stage earlier than ATT, indicating a faster onset of effective inhibition and a comparatively stronger initial inhibition response. Overall, all three proteins exhibit obvious inhibition activity compared to the control, with the inhibition ability follows the order AAT > ATAT > ATT, consistent with prior experimental observations.[38] This agreement validates the reliability of our model system and confirming that subtle variations in residue composition can strongly modulate ice inhibition activity.

To elucidate the physical origin of their differing ice inhibition abilities, we examined their binding behaviors at the ice-water interface. Representative snapshots from the ice inhibition simulations show that all three proteins bind to the ice surface during the inhibition process, suppressing the growth of ice (Figure 3a). However, they exhibit varying degrees of "upright" orientation relative to the ice plane. To quantify this, we measured the cosine of the angle ($\cos\theta$) between the peptide backbone vector and the z-axis (perpendicular to the ice plane), as shown in Figure 3b. AAT predominantly adopts a binding configuration nearly parallel to the ice surface ($\theta > 80°$), forming a strong binding state. In contrast, ATAT and ATT tend to rotate upon adsorption, with one end of the backbone lifting away from the ice surface, resulting in a smaller $\theta$ (typically $< 65°$) and relatively weak ice-binding interactions. Furthermore, we also calculated the distance between the proteins and the ice surface (Figure 3c). The average distances follow the order AAT (0.8 nm) < ATAT ≈ ATT (1.3 nm), with larger fluctuations observed for proteins with higher T* content. These results collectively indicate that the proteins exhibit distinct ice-binding behaviors, with AAT forming more stable and tightly bound configurations, whereas ATAT and ATT predominantly occupy weaker binding states.

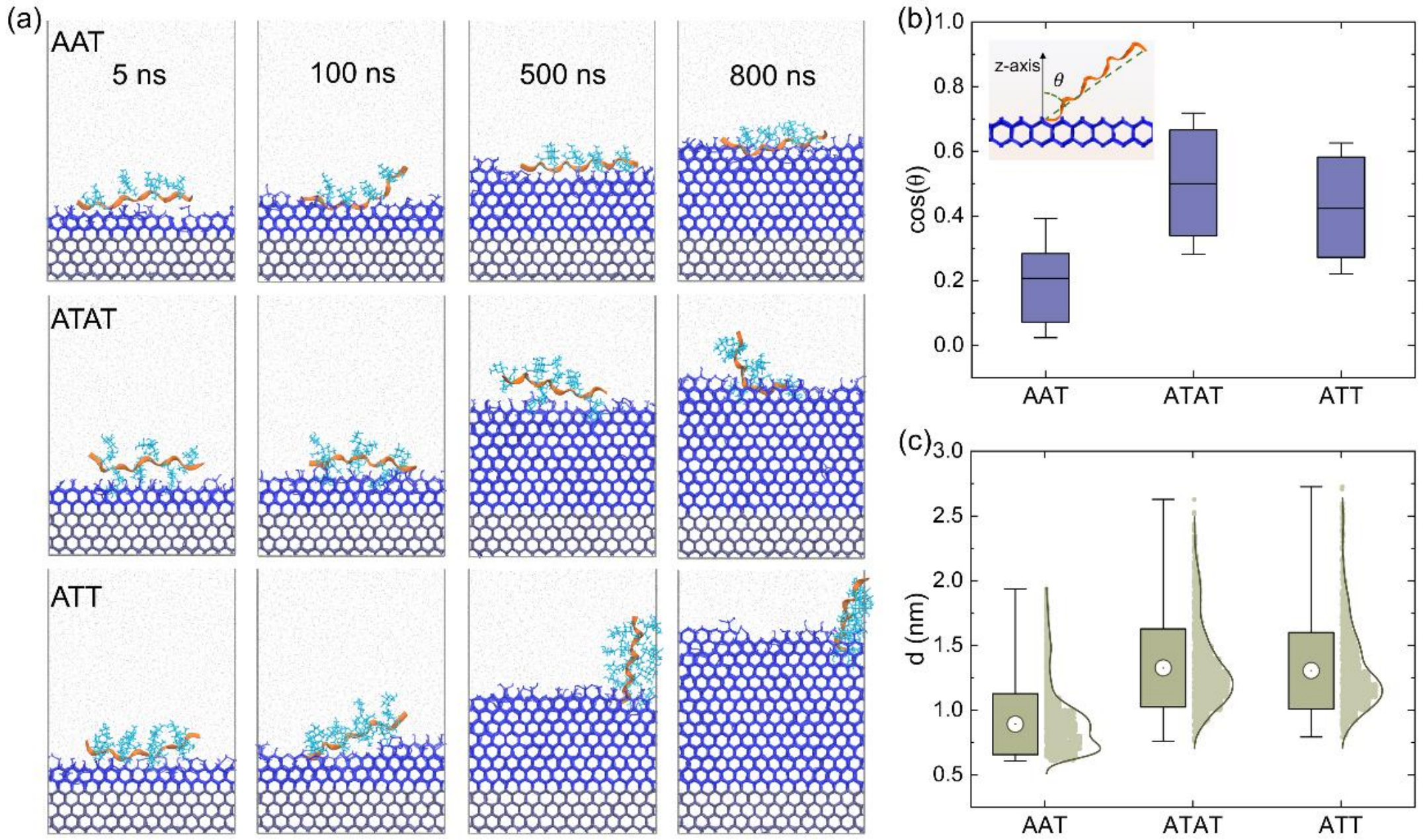

Figure 3. Binding behavior of the proteins at the ice surface. (a) Representative simulation snapshots showing ice growth in the presence of AAT, ATAT, and ATT. The bottom four ice layers (ice blue) are restrained by a harmonic potential. Free ice is depicted as blue wireframes, and liquid water molecules are shown as gray dots. Peptides are color-coded as in Figure 1c. (b) Distribution of the cosine of the angle between the peptide backbone vector and the z-axis during ice growth inhibition simulations, reflecting alignment with the ice surface. (c) Distribution of the vertical distance between the protein center of mass and the underlying ice surface directly beneath the protein.

For a deeper mechanistic understanding of their differing ice-binding abilities, we performed well-tempered metadynamics simulations to compute the binding free energy on the ice surface (see Methods for details). The resulting two-dimensional free energy landscapes (Figure 4a–c) are projected as functions of the peptide-ice distance (d) and backbone orientation ($\cos\theta$, where $\cos\theta \approx 0$ denotes a configuration nearly parallel to the ice plane). The binding states identified in ice inhibition simulations (Figure 2) are localized within black dashed boxes on these free energy landscapes. Notably, AAT accesses a global free energy minimum located at d ≈ 0.8 nm and $\cos\theta < 0.2$, indicative of a thermodynamically stable and strong binding state characterized by parallel alignment and direct contact with the ice surface. In contrast, ATAT and ATT predominantly sample the higher-energy states featuring larger protein-ice distances (~1.2 nm) and more tilted backbone orientations ($\cos\theta > 0.4$), corresponding to thermodynamically metastable and weaker binding states. These results suggest that AAT achieves a substantially more favorable ice-binding interaction upon contacting with ice, leading to stronger and more stable adhesion compared to ATAT and ATT.

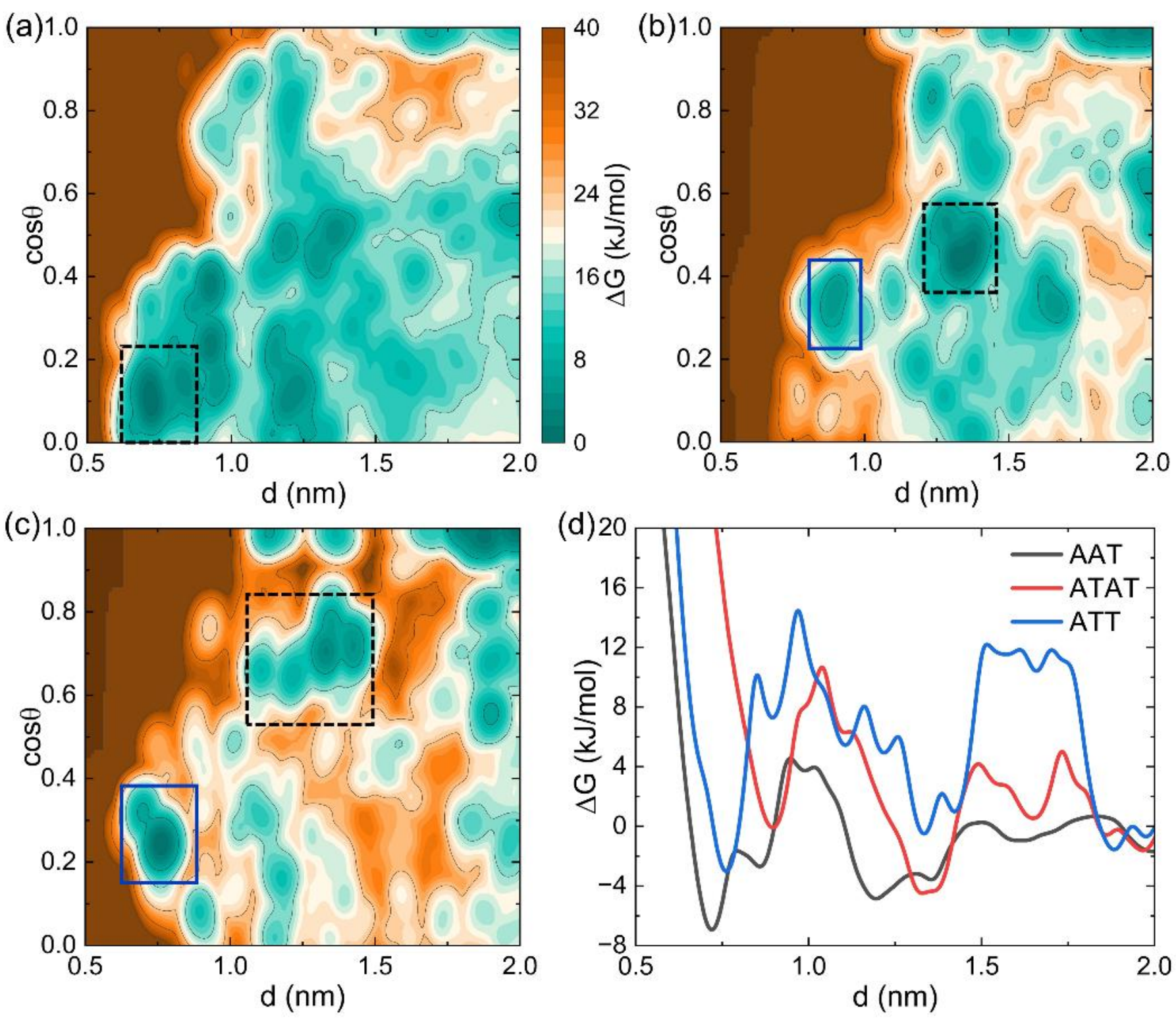

Figure 4. Binding free energy landscapes of the proteins on the ice prismatic plane. (a-c) Two-dimensional free energy profiles for AAT (a), ATAT (b), and ATT (c) on top of ice surfaces.

Binding states consistent with those observed in the ice growth inhibition simulations (Figure 2) are outlined by black dashed rectangles. Additional low-energy binding states that are thermodynamically favorable but rarely accessed by AAT and ATT during the ice inhibition simulations are highlighted by blue solid rectangles. (d) One-dimensional binding free energy profiles of the proteins as a function of the distance between the protein center of mass and the prismatic ice plane along the z-axis. The baseline (i.e., the zero of the free energy profiles) is set by the average free energy in the pre-binding region (1.8-2.0 nm). All data are obtained from WTMetaD simulations, where two CVs are employed to bias: the distance (d) between the protein center of mass and the ice surface, and the cosine of the angle (cos θ) between the protein backbone and the z-axis.

More importantly, although both ATAT and ATT possess thermodynamically stable binding states near the ice surface (blue boxes), these states remain kinetically inaccessible during the timescales of the ice inhibition simulations. To elucidate this phenomenon, we constructed one-dimensional free energy profiles as functions of protein–ice distance (Figure 4d). All three proteins exhibit a pronounced free energy minimum at d ≈ 0.7~0.9 nm, confirming the existence of thermodynamically stable strong-binding states achieved upon direct contact with the ice surface. However, the free energy barrier for adsorption—defined as the larger of the two local energy maxima along the pathway from the unbound state near the ice-water interface (d ≈ 2.0 nm) to the strong binding state (d < 1.0 nm)—increases markedly with rising T* content: ATT > ATAT > AAT. Relative to AAT, the barriers elevate by ~7 and ~10 kJ/mol for ATAT and ATT, respectively. The corresponding adsorption rate constants can be estimated using the Arrhenius-like relation:

$$k = A \exp\left(-\frac{\Delta G^{\ddagger}}{k_B T}\right) \tag{1}$$

where $A$ is a pre-exponential factor determined by the structural characteristics of the antifreeze protein, $\Delta G^{\ddagger}$ is the adsorption barrier, $k_B$ is the Boltzmann constant, and $T$ is the temperature. Assuming comparable values of $A$ for all three glycoproteins and considering $T = 268$ K, the relative adsorption efficiency of ATAT and ATT is reduced to approximately 4.3% and 1.1%, respectively, compared to AAT. While these numerical estimates may not be exact, the qualitative trend is clear: the kinetic barriers increase with T* content, reducing the accessibility of the strong ice-binding states. This trend is consistent with our ice inhibition simulations (Figure 2), which show that ATAT and ATT require significantly longer times to effectively adsorb to the ice surface and achieve substantial suppression of ice growth. These results collectively indicate that the elevated kinetic barriers substantially hinder the formation of strong ice-binding states for ATAT and ATT, thereby limiting their adsorption and accumulation at the ice interface.

In conjunction with previous fluorescence experiments demonstrating that antifreeze molecules occupying strong ice-binding states dominate the suppression of advancing ice fronts,[40] we conclude that the diminished ice inhibition activity of ATAT and ATT arises primarily from their decreased accessibility of the strong binding state required for potent IRI efficiency.

To validated this hypothesis, we performed additional ice growth inhibition simulations in which the proteins were initially restrained to achieve a tightly bound state. Once this configuration was established, all restraints were removed to assess their intrinsic inhibitory performance under the same binding conditions. As shown in Figure 5, achieve stable binding to the ice surface and display comparable inhibition activity once the strong binding state is established. This further demonstrates that the decline in IRI activity with increasing T* content originates primarily from their intrinsic capacity to attain and maintain the tight ice-binding state, rather than an absence of such states. Therefore, we conclude that the kinetic accessibility of the strong binding state, governed by the adsorption barrier, plays a key role in dictating their IRI activity.

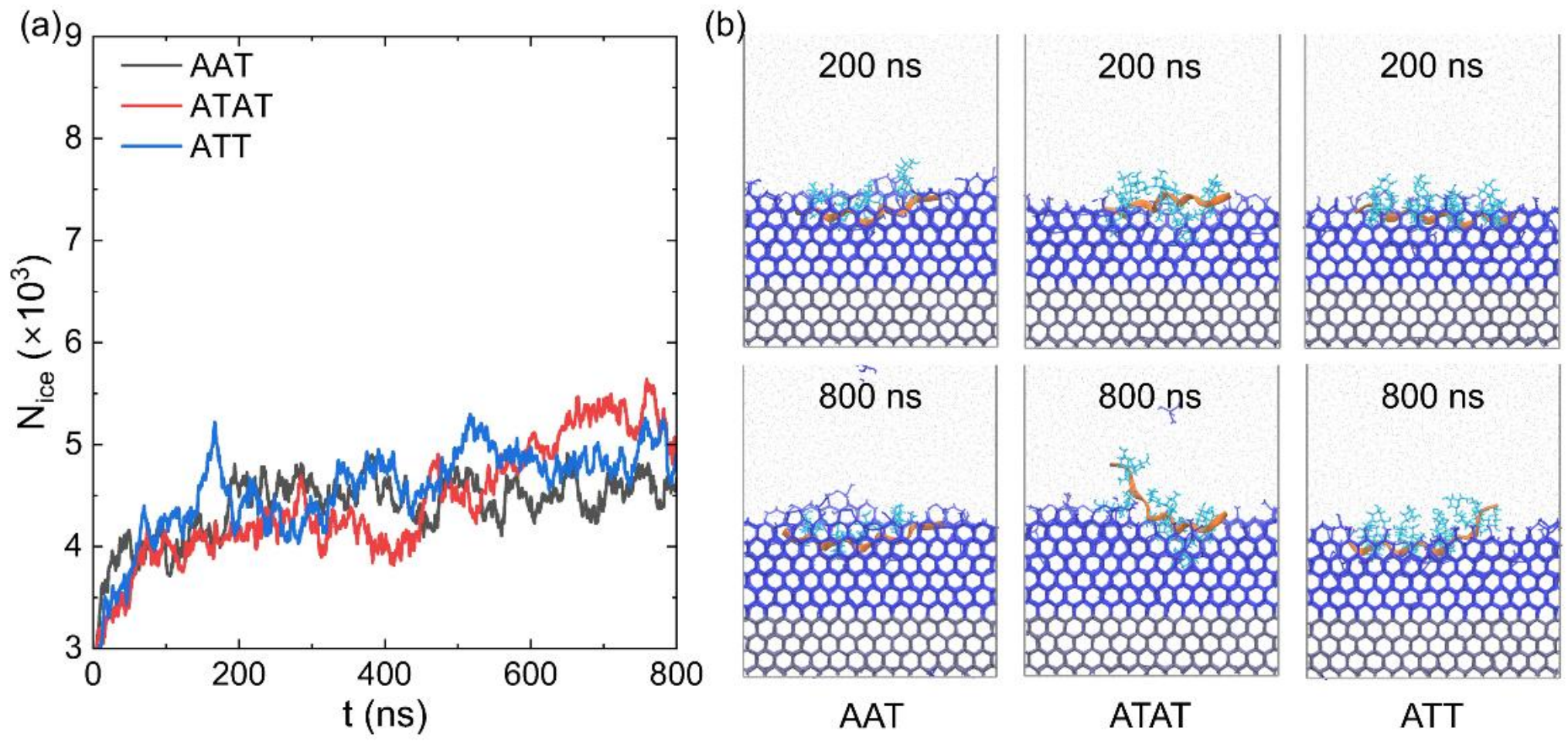

Figure 5. All three proteins are capable of forming strong ice-binding states, leading to enhanced ice growth inhibition. (a) Time evolution of the number of ice-like water molecules identified by the CHILL+ algorithm[60] during pre-binding ice inhibition simulations. (b) Representative snapshots from these simulations. The color scheme follows that of Figure 3a. In these pre-binding ice inhibition simulations, harmonic restraints were initially applied to the protein backbone to maintain a strong-binding configuration with the ice surface for the first 200 ns. Afterward, the restraints were released to examine how the tightly bound proteins influence ice growth.

Building upon the importance of kinetic accessibility, we next investigate how increasing T* content modulates adsorption kinetics. Under strong-binding conditions, glycan moieties directly anchor to the ice surface and thereby enabling the strong binding (Figure 5b), consistent with previous simulation studies.[33,34] However, their high hydrophilicity necessitates partial desolvation during adsorption, imposing a thermodynamic penalty due to partial disruption of hydrogen bonds between the protein and surrounding water molecules that manifests as an elevated adsorption barrier. As noted in studies of other antifreeze proteins,[73] excessively strong enthalpic interactions with surrounding water can paradoxically hinder adsorption because the hydrogen bonds between the protein's polar residues and water molecules must be broken to allow binding with the ice surface. This disruption represents an enthalpic cost, which slows down the adsorption process. Consequently, increasing T* content raises the adsorption barrier, impeding the adsorption kinetics. In contrast, Ala-associated methyl groups compensate entropically by releasing ordered hydration water,[26] facilitating ice association despite weaker inherent binding strength. Thus, while moderate T* content enables effective ice anchoring, excessive glycan moieties raise the adsorption barrier beyond thresholds, limiting their effective accumulation and accessibility of the strong binding states at the ice surface.

In light of the above results, it is important to characterize the key structural features governing ice-binding behavior, enabling structure-based design of related antifreeze materials. Extensive evidence demonstrated that spatial segregation of hydrophilic and hydrophobic moieties—commonly referred as facial amphiphilicity—play an important role in the binding of antifreeze agents to ice.[12,13,16,32,68,74-76] To quantify such a segregation in these proteins, we employed a vector-based approach adapted from Mallajosyula et al.,[69] which quantifies the collective orientation of glycan chains relative to the backbone. Specifically, we computed the net orientation vector $s$, defined as:

$$\theta_1 = 0, \theta_{i+1} = \theta_i + \phi_i \Rightarrow \theta_{i+1} = \sum_{j=1}^{i} \phi_j \tag{2}$$

$$s = \frac{1}{N}\left(\sqrt[2]{s_x^2 + s_y^2}\right), \text{where } s_x = \sum_{i=1}^{N} \cos(\theta_i), s_y = \sum_{i=1}^{N} \sin(\theta_i) \tag{3}$$

Here, $N$ is the total number of T* residues, $\phi_i$ is the pseudo-dihedral angle between the β-oxygen of the disaccharide and the Cα atom of the corresponding T*. Higher $s$ value (approaching 1) indicates greater directional alignment of the glycans on one face of the peptide, reflecting enhanced facial amphiphilicity.

As shown in Figure 6a, the distributions of $s$ values derived from the REMD trajectories exhibit distinct peaks of each protein (Figure 1c), located at 0.95 (AAT), 0.18 (ATAT), and 0.38 (ATT), respectively. This suggest that glycan orientation order follows the trend AAT > ATT > ATAT. This trend correlates well with the thermodynamic stability of the tight ice-binding states observed in the free energy profiles (Figure 4d). In view of previous evidence that ordered glycan orientations promote spatial complementarity with the ice surface through distance matching,[68] this parament can capture the structural basis for ice-binding stability to some extent. However, this trend contradicts their ice binding affinities and IRI activities (ATAT > ATT), implying spatial segregation alone cannot explain their functional divergence.

Notably, recent experiments have shown that synthetic AFGPs can retain high IRI activity as long as the Ala:T* ratio remains at 2:1, even when the sequence order is scrambled.[77] These results highlight the importance of considering both the spatial organization and relative abundance of hydrophilic and hydrophobic residues, as these jointly influence ice-binding efficiency by modulating both binding stability and adsorption barrier.

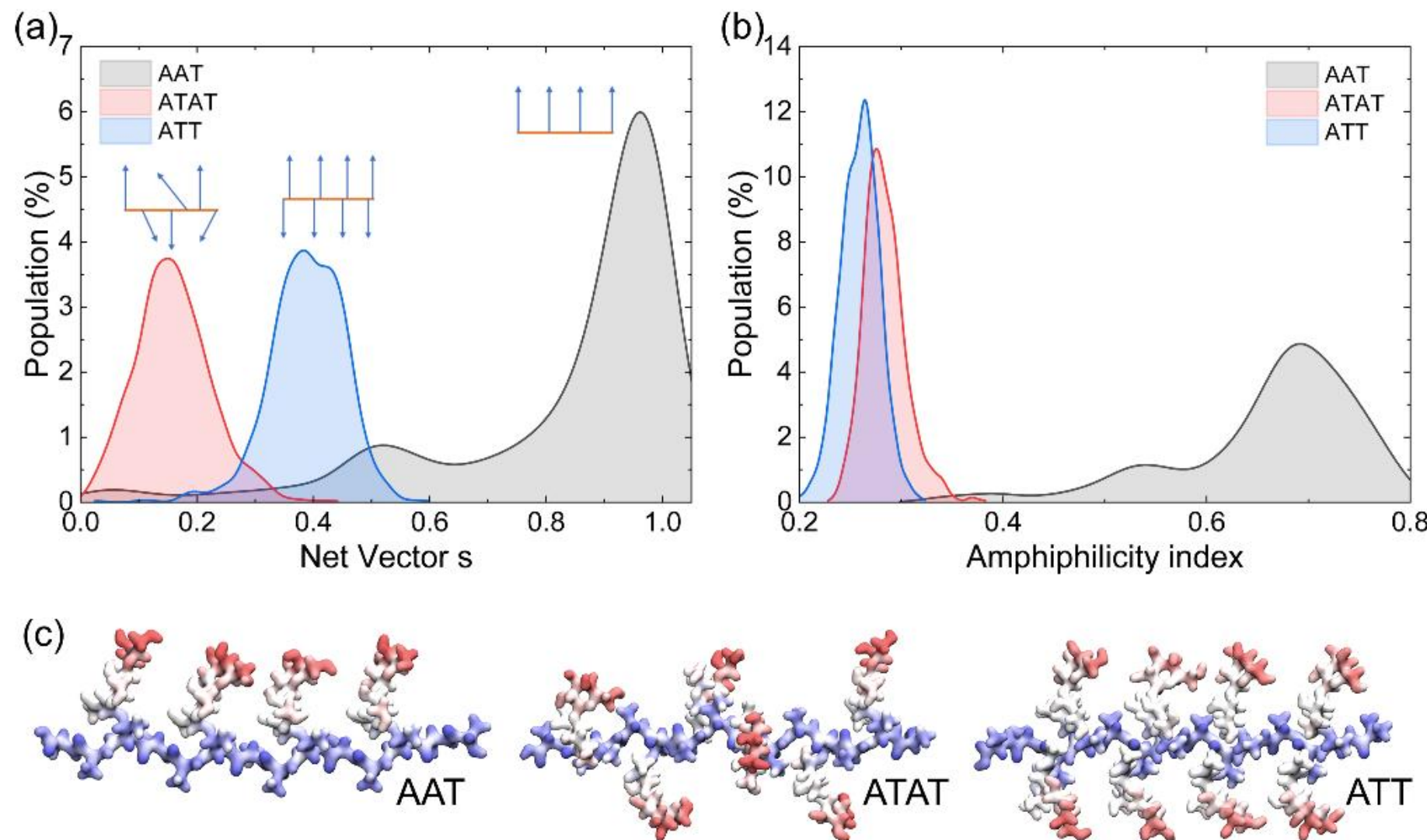

Figure 6. Amphiphilicity analysis of proteins. (a) Magnitude of the net amphiphilicity vector (s), calculated according to Equation 2. (b) Amphiphilicity index derived from the SASA of hydrophobic and hydrophilic regions, as defined in Equation 3. (c) Spatial orientation of glycan moieties in AAT, ATAT, and ATT. Color coding is based on the distance from the backbone: the backbone is shown in blue, and glycan residues are shown in red.

To reconcile these effects, we developed a composite facial amphiphilicity index (FAI) that integrates both the glycan orientation (quantified by the net vector $s$) and the relative abundance of T* and Ala residues (estimated via their solvent-accessible surface areas). The FAI is defined as:

$$FAI = \frac{\left[s\left(1 - \frac{|SASA_{T^*} - 0.5 \cdot SASA_{tot}|}{0.5 \cdot SASA_{tot}}\right) + \left(1 - \frac{|SASA_{Ala} - 0.5 \cdot SASA_{tot}|}{0.5 \cdot SASA_{tot}}\right)\right]}{2} \quad (4)$$

Here, $SASA_{T*}$, $SASA_{Ala}$, and $SASA_{tot}$ denote the solvent-accessible surface areas of T* residues, Ala residues, and the entire molecule, respectively. This normalized parameter achieves maximal value when the spatial distribution of glycans is highly directional ($s = 1$) to and the hydrophilic/hydrophobic components are well-balanced. Physically, disordered glycan chains can disrupt the hydrophobic/hydrophilic balance by sterically shielding hydrophobic residues, thereby affecting ice-binding efficiency. Thus, this parameter captures the combined effects of surface hydrophobicity and glycan chain orientation, which together influence the spatial complementarity and adsorption behavior of the protein on the ice surface.

As shown in Figure 6b, AAT exhibits the largest FAI value, followed by ATAT and ATT, consistent with their respective ice binding affinities and IRI activities. Notably, although ATT displays a more ordered glycan orientation than ATAT (Figure 6c), its overall amphiphilicity is lower due to its the overrepresentation of hydrophilic residues. These results indicate that optimal antifreeze activity arises from a synergistic effect between directional glycan orientation and hydrophobic-hydrophilic composition.

Collectively, these findings establish the FAI as a minimal yet robust descriptor that captures the key structural features that correlate strongly with ice-binding ability. We therefore propose that maximizing the FAI, by fine-tuning both sequence composition and three-dimensional spatial organization of the molecule, provides a rational design strategy for engineering potent antifreeze proteins and synthetic mimetics. Looking ahead, further refinements to the FAI framework, such as the incorporation of hydrophobic moments[31] or the adoption of residue-specific hydrophobicity scales[78-80], will further enable quantitative understanding of the relationships between molecular amphiphilicity and IRI activity.

## Conclusions

In this work, we employed all-atom MD simulations to systematically investigate the structural and mechanistic origins of IRI activity of proteins with increasing T* content (AAT, ATAT, and ATT). While all three peptides adopting PPII-like backbone conformations, their distinct glycan spatial distributions yield markedly different dominant conformations and ice-binding behaviors. Crucially, our simulations reveal that the experimentally observed decline in IRI efficiency with increasing T* content stems not from a loss of strong ice-binding states, but from the diminished kinetic accessibility of these states during adsorption. Mechanistically, hydrophilic glycan moieties enhance binding stability via direct hydrogen-bond anchoring to the ice surface yet impose substantial desolvation penalties due to their strong hydration property. This manifests as an elevated free energy barrier during adsorption, significantly retarding the adsorption kinetics and ultimately diminishing IRI efficacy. Consequently, excessive T* content heightens this adsorption barrier, reducing the ice-binding efficiency and directly accounting for the observed IRI trend: AAT > ATAT > ATT.[38]

Building upon this mechanistic insight, we propose a facial amphiphilicity index (FAI) that integrates glycan orientation order and molecular compositions quantitatively characterize the structural basis underlying their activity. The observed correlation between FAI and IRI activity directly bridges the microscopic amphiphilic architecture with the macroscopic antifreeze performance, highlighting the importance of jointly optimizing residue composition and spatial patterning to maximize ice-binding affinity and IRI efficacy. Collectively, this work advances the structure-activity understanding of antifreeze proteins and establishes facial amphiphilicity engineering as a rational design paradigm for next-generation IRI materials.

## AUTHOR INFORMATION

### Corresponding Author

* Zhaoru Sun—School of Physical Science and Technology, ShanghaiTech University, Shanghai 201210, China.

Email: sunzhr@shanghaitech.edu.cn

### Notes

The authors declare no competing financial interest.

## ACKNOWLEDGMENT

This work was supported by the Shanghai Rising-Star Program (23QA1406800). The authors also thank the computing resources and technical support provided by the High-Performance Computing (HPC) Platform of ShanghaiTech University and State Key Laboratory of Molecular Engineering of Polymers (Fudan University).